\documentclass[aps,twocolumn,showpacs,superscriptaddress,reprint,prb]{revtex4-2}

\usepackage{bm}
\usepackage{graphicx}
\usepackage{amsmath}
\usepackage{amsfonts}
\usepackage{siunitx}
\usepackage{bbold}
\usepackage{color}
\usepackage{comment}
\usepackage{textcomp}
\usepackage{gensymb}
\usepackage{soul}
\usepackage[T1]{fontenc} 
\providecommand{\abs}[1]{\lvert#1\rvert}

\begin{document}

\title{Tuning magnetism in graphene nanoribbons via strain and adatoms}

\author{Pablo Moles}
\email{pmoles@ucm.es}
\affiliation{GISC, Departamento de F\'{\i}sica de Materiales, Facultad de Ciencias Físicas, Universidad Complutense, E-28040 Madrid, Spain}

\author{Hernán Santos}
\email{Hernan.Santos@uclm.es}
\affiliation{Departamento de Física Aplicada, Facultad de Ciencias Ambientales y Bioquímica, Universidad de Castilla-La Mancha, Campus Tecnológico de la Fábrica de Armas, Avenida Carlos III s/n, 45071 Toledo, Spain}
\affiliation{Departamento de Matem\'atica Aplicada, Ciencia e Ingenier\'{\i}a de Materiales y Tecnolog\'{\i}a Electr\'onica, ESCET, Universidad Rey Juan Carlos, C/ Tulip\'an s/n, M\'ostoles 28933, Madrid, Spain}

\author{Francisco Dom\'{\i}nguez-Adame}
\affiliation{GISC, Departamento de F\'{\i}sica de Materiales, Facultad de Ciencias Físicas, Universidad Complutense, E-28040 Madrid, Spain}

\author{Leonor Chico}
\affiliation{GISC, Departamento de F\'{\i}sica de Materiales, Facultad de Ciencias Físicas, Universidad Complutense, E-28040 Madrid, Spain}



\begin{abstract}


We investigate the impact of strain and adsorbed H adatoms on the magnetic properties of zigzag graphene nanoribbons (ZGNRs) using a combination of tight-binding and density functional theory methods for both, ferromagnetic (FM) and antiferromagnetic edge configurations (AFM). Our study reveals that longitudinal strain induces a significant enhancement in the edge magnetic moment, that we attribute to strain-driven modifications in the band structure. 
In addition, we describe H~adatoms within the tight-binding approach by employing both an unrelaxed vacancy model and the Anderson impurity model. By comparing to density functional theory results, we corroborate that the Anderson impurity model is best suited to describe H adsorption. We then focus on the metallic FM edge configuration of the ZGNRs to better exploit the tuning of its properties. We find that the magnetic configuration of H~adatoms is strongly influenced by the edges, with an AFM coupling between edges and the H~adatom. In fact, the magnetic spatial pattern of the H adatom differs to that found in graphene due to this edge coupling. Importantly, we find robust discrete plateaus of integer magnetic moment as strain is varied in the defected ZGNRs, that we relate to changes in the band structure, namely, a half-metallic character or the opening of a gap. This behavior can be of interest for magnetic applications of carbon-based nanostructures. 
\end{abstract}

\maketitle

\section{Introduction}


Partially filled $d$ and $f$ orbitals are responsible of the magnetic order of transition metals and rare earth ions in common magnetic materials. However, their  environmentally harmful nature has boosted the quest of alternative routes to achieve magnetic order in solids.
%
%
In this context, $p$--orbital magnetism has arisen in the last two decades as a possible way to realize novel and sustainable spintronic devices~\cite{Volnianska2010}. 
It can emerge in low-dimensional carbon-based materials such as nanostructured graphene, nanoflakes or nanoribbons. At the edges or vacancies of such nanomaterials, the $p_z$ orbitals of carbon atoms give rise to $\pi$-electronic states, in which electron-electron interactions induce magnetic ordering, referred to as $\pi$-magnetism~\cite{Yazyev2010review,review2022platform}.

Magnetic graphene nanostructures are particularly promising for spintronics~\cite{colloquium2020spintronics,han2014graphenespintronics}. Graphene exhibits weak spin-orbit and hyperfine couplings~\cite{yazyev2008hyperfine,min2006SOC,huertas2006SOC}, which are the main physical mechanisms for relaxation and decoherence of electron spins. This, in addition to the high electron mobility in this material, results in the longest spin diffusion length achieved at room temperature~\cite{drogeler2016mobility}. These characteristics are fundamental for the application of graphene-based materials in spintronic devices. 

Bulk pristine graphene is intrinsically non-magnetic. However, there are two main scenarios in which magnetism may emerge in this material. The first one is related to the presence of point defects in the graphene lattice, specifically vacancies and adatoms~\cite{Yazyev2010review}.
This type of magnetism has been extensively studied~\cite{yazyev2007,yazyev2008,rossier2008armchair}, and  experimentally observed in irradiated graphite samples~\cite{FMOhldag2007,FMBarzola2007,Brihuega2010}. While these early studies lacked precision over the distribution of defects, advancements in experimental techniques have allowed for the controlled creation of defects. The adsorption of adatoms, particularly hydrogen, represents one of the most effective methods for modulating the magnetic properties in graphene with atomic precision~\cite{brihuega2016}.
The second scenario involves certain graphene nanostructures, namely graphene nanoflakes~\cite{Yazyev2010review} and zigzag graphene nanoribbons (ZGNRs), which present intrinsic magnetic ordering without the need of other defects than the edges.  
As it is well-known, these nanoribbons develop low-energy states localized at the zigzag edges, which are predicted to be spin-polarized due to electron-electron interactions~\cite{Pisani2007magneticribbons, Palacios2010nanoribbons,Hernan2007Unzipping}. ZGNRs can present half-metallicity, producing fully spin-polarized currents~\cite{son2006halfmetallic}. Moreover, magnetoresistive devices~\cite{Muñoz-Rojasmagnetoresistive}, spin valves~\cite{kim2008spin-valve}, spin diodes~\cite{spin-diode} and field-effect transistors~\cite{Vancsó2017FET} based in ZGNRs have been proposed to exploit their magnetic edge states, constituting the building blocks of prospective graphene-based spintronics. 

The lack of atomic precision in early synthesis methods of ZGNRs, such as solution-phase chemistry~\cite{2008chemically} or top-down approaches~\cite{2009nanowiremask}, hampered the experimental verification of magnetism in these systems, mainly because magnetic ordering is very sensitive to edge roughness~\cite{Huang2008}. However, the development of on-surface-synthesis techniques for zigzag nanoribbons has provided the observation of energy gaps and local density of states consistent with the existence of edge magnetism~\cite{ruffieux2016on-surfaceZGNR}. Although the direct experimental evidence of magnetism in ZGNRs still remains elusive~\cite{detection2023}, the atomic precision achieved both in synthesis and in defect creation in graphene nanostructures is a powerful motivation to explore the interplay of these two sources of magnetism in ZGNRs and the possible ways to modify it~\cite{kaxiras}.

In this work, we propose a way to selectively enhance the magnetic response of ZGNRs by different means. Firstly, we analyze the role of induced strain as a way to tune their electronic properties~\cite{2021straintronics,Naumis_2017strain}. Using first-principles and tight-binding~(TB) methods, we consider uniaxial strain along the zigzag direction of a ZGNR, resulting in a smooth, albeit important, enhancement of its magnetic response. We attribute this increase to modifications in the band structure produced by the strain field. Furthermore, we also explore the tunability of the magnetic response via the presence of point defects, namely H~atoms adsorbed in the ZNGR, focusing on the ferromagnetic edge configuration, which we find more adequate for such property tuning. 
We find that the magnetic moment varies with strain yielding robust discrete plateaus of integer magnetic moment in the defected ZGNRs. These plateaus can be explained by resorting to the band structure, being related to the half-metallic character of the gap opening with strain, which may be relevant for magnetic applications of graphene nanoribbons. 

The article is organized as follows. Section~\ref{sec:Model} describes the system under consideration and outlines the computational methods employed. Specifically, we detail the tight-binding model in section~\ref{sec:Model}.A and the density functional theory (DFT) calculations in section~\ref{sec:Model}.B. Our results are presented in section \ref{sec:results}, divided into three main parts. Firstly, section~\ref{sec:results}.A investigates the effects of strain on pristine ZGNRs. Secondly, section~\ref{sec:results}.B examines the adsorption of H adatoms, and finally, section~\ref{sec:results}.C studies the role H adatoms in combination with strain on ZGNRs. Section~IV concludes with a brief summary of our main findings.

\section{System and Computational Methods}\label{sec:Model}

\begin{figure}[ht]
    \centering
    \includegraphics[width=0.95\columnwidth]{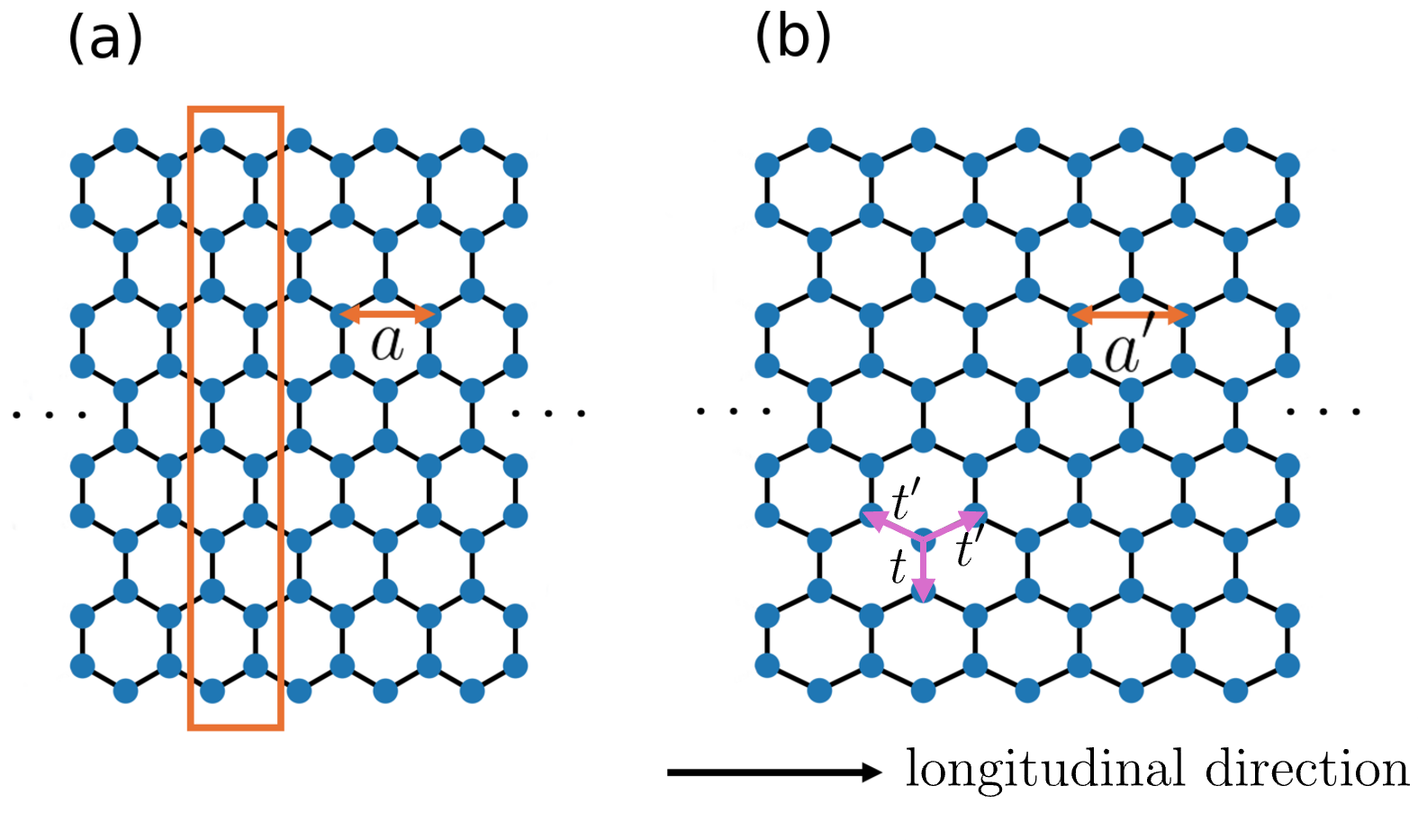}
    \caption{(a) Atomic arrangement of an unstrained 8-ZGNR. The unit cell and the lattice parameter are displayed. (b)~The same ZGNR with induced longitudinal strain. The strain-modified lattice and hopping parameters are also shown.}
    \label{fig1}
\end{figure}

The system under consideration consists of an infinitely long ZGNR, with a lattice parameter $a=\sqrt{3}\,a_{0}$, being $a_{0}=1.42$\,\AA\ the C--C distance. We label the nanoribbon according to its width, $W$-ZGNR, where $W$ indicates the number of zigzag chains of atoms across the width of the nanoribbon. Thus, the nanoribbon has $N=2W$ C~atoms in its translational unit cell [see Fig.~\ref{fig1}(a) for further details].

We consider that a uniform and uniaxial strain is applied along the zigzag direction of the nanoribbon. The crystal structure of the nanoribbon is modified in this direction, altering the interatomic distances and thus the lattice parameter, denoted as $a'$. As further discussed later, we assume that the width of the ZNGR remains unchanged after uniaxial stress. Figure~\ref{fig1}(b) shows this structural modification.
%
%
For a ZGNR with this applied uniaxial strain, the size change corresponds directly to the modification of the lattice parameter in that direction. Thus, we quantify the strain as 
\begin{equation}
\epsilon=\frac{a'-a}{a}\ .
\label{strain}
\end{equation}
Admittedly, this assumption does not accurately describe the atomic rearrangement in a real situation, where the ZGNR would be also relaxed in the transverse direction due to the Poisson effect~\cite{Naumis_2017strain}. However, the Poisson effect is relatively small in graphene, with Poisson ratios ranging from $0.14$ to $0.19$~\cite{poissonratio, poissonsanchez-portal}. Furthermore, we have performed DFT calculations that account for Poisson relaxation (see details in section~\ref{sec:Model}.B) and found that it does not substantially alter the magnetic response [see Fig.~\ref{fig2}(b)]. Consequently, we have chosen to focus on the simplified description, which aims to clarify the main underlying physics related to magnetism.

We also consider H~adatoms adsorbed onto the ZGNR. It is well-established that H~adatoms form covalent bonds with C~atoms, where the $1s$ H orbital hybridizes with the $2p_{z}$ C orbital. 
%
%
However, the presence of H~adatoms does not lead to significant deformation of the graphene lattice. It only induces a slight out-of-plane relaxation in the surrounding region~\cite{yazyev2007,Yazyev2010review}. Our DFT calculations confirm that this relaxation does not result in noticeable changes of the electronic structure. Consequently, in this study, we neglect the out-of-plane relaxation when modeling the occurrence of H~adatoms and focus on the magnetic changes induced in the system.



\subsection{Tight-binding model}

The TB approach followed in this work employs a one-orbital mean-field Hubbard model, a well-known and extensively used approximation to describe magnetism in carbon materials~\cite{Yazyev2010review}. Its results have been demonstrated to be in good agreement with first-principles calculations~\cite{Rossier2007nanoislands,gunlycke2007graphene,Rossier2008multiferroic,soriano2010hydrogenated}. The Hubbard model only considers  the non-hybridized $p_{z}$ orbital of each C~atom, which contributes with one electron to the resulting $\pi$ band. For undoped or unbiased 
graphene that we considered, 
the ZGNR is half-filled. The Hamiltonian of the system  splits into two parts, $\mathcal{H}=\mathcal{H}_{0}+\mathcal{H}_{U}$, where
$\mathcal{H}_{0}$ corresponds to the non-interacting TB Hamiltonian,  and 
$\mathcal{H}_{U}$ represents the electron-electron interaction. 
Here, the non-interacting TB Hamiltonian only includes nearest-neighbor couplings, 
\begin{equation}
    \mathcal{H}_{0}=-\sum_{\langle i,j\rangle,\sigma}t'\,c^\dagger_{ i,\sigma}\,c_{j,\sigma}^{}+\mathrm{H.c.}
    \label{H0}
\end{equation}
where $\mathrm{H.c.}$ stands for Hermitian conjugate and the summation in $\langle i,j\rangle$ runs over nearest-neighbor C~atoms. The origin of energy is set at the energy of the $p_z$ orbital. Here $c^{\dagger}_{i}$ and $c_{j}^{}$ are the creation and annihilation fermion operators at atoms $i$ and $j$, respectively. The parameter $t'$ denotes the nearest-neighbor hopping energy and $\sigma=\uparrow,\downarrow$ indicates the electron spin. 

In a TB model, the hopping parameter depends on the interatomic distance, in this case, is altered due to the strain. Among the three possible first-neighbor hoppings, the transverse one remains unchanged, since the distance between atoms remains constant in this direction. However, the other two hoppings between bonds with a longitudinal component are modified due to the variation on the atomic separation in the direction of strain. We denote these as $t'$ [see Fig.\ref{fig1}(b)]. We account for this effect by modifying the hopping parameters with the distance according to the expression~\cite{parametrizacion}
\begin{equation}
    t'=t\,e^{-\beta(r/a_{0}-1)}\ ,
    \label{hopping}
\end{equation}
where $t=2.5$\,eV is the hopping in the strain-free case, $r$ is the distance between the atoms and $\beta=3.1$. 
%
%
The parametrization~\eqref{hopping} has been successfully validated with DFT results~\cite{ribeiro2009strained}.

The magnetic response of the ZGNR is modeled with the second term of the Hamiltonian, $\mathcal{H}_{U}$. The Hubbard model introduces these interactions by means of an onsite Coulomb repulsion. Thus, electrons with opposite spin occupying the same site experience a repulsion quantified by the energy $U>0$, known as the Hubbard parameter. The interaction Hamiltonian is 
\begin{equation}
  \mathcal{H}_{U}=U\sum_{i}n_{i,\uparrow}n_{i,\downarrow}\ ,     
\end{equation}
where $n_{i,\sigma}=c^\dagger_{ i,\sigma}\,c_{i,\sigma}^{}$ is the number operator, which gives the spin-resolved electron density at atom $i$. We set the value of $U=3$\,eV in our numerical calculations. To deal with the many-body interaction Hamiltonian $\mathcal{H}_{U}$, the restricted Hartree-Fock mean-field approximation is adopted. The resulting Hamiltonian is approximated as
\begin{equation}
    \mathcal{H}_{U}=U\sum_{i}\big(n_{i,\uparrow}\langle n_{i,\downarrow}\rangle+n_{i,\downarrow}\langle n_{i,\uparrow}\rangle-\langle n_{i,\uparrow}\rangle\langle n_{i,\downarrow}\rangle\big)\ .
\end{equation}
Here, the spin-up and spin-down matrix elements at site $i$ depend on $\langle n_{i,\downarrow}\rangle$ and $\langle n_{i,\uparrow}\rangle$ respectively, which represent the average electron population 
with opposite spins at that site. These numbers are the expectation values of the spin-resolved electron densities obtained from the eigenvectors of $\mathcal{H}$, which are initially unknown. The Hamiltonian is solved self-consistently, using the following procedure: starting from an initial guess of $\langle n_{i,\sigma}\rangle$ chosen randomly, (i) the $2N\times2N$ matrix representation of the Hamiltonian is obtained. (ii) The Hamiltonian $\mathcal{H}(k)$ is diagonalized in the reciprocal space, and the corresponding spin-polarized eigenvectors $\phi_{v,i,\sigma}(k)$ are computed, where $v$ is the band index. (iii) The updated spin-resolved densities are obtained as follows: 
\begin{equation}
     \langle n_{i,\sigma}\rangle=\frac{1}{2\pi}\sum_{v=1}^{N}\int_{-\pi/a}^{\pi/a}\phi^{\dagger}_{v,i,\sigma}(k)\phi_{v,i,\sigma}(k)\,\mathrm{d}k\ ,
     \label{ocupacion}
\end{equation}
where the summation runs up to the $N$th--band since the system is half-filled. We employ a fine grid of $5000$  $k$ wavenumbers over the Brillouin zone to perform the numerical integration in equation~(\ref{ocupacion}). Given the new $\langle n_{i,\sigma}\rangle$ values, the steps (i)--(iii) are repeated iteratively until convergence is reached for the values of the electron densities. A more detailed explanation can be found in reference~\cite{yuriko} for a similar self-consistent algorithm.

From the imbalance of the obtained spin-polarized densities of electrons, a local site-resolved magnetic moment arises
\begin{equation}
     m_{i}=\mu_{B}\big(\langle n_{i,\uparrow}\rangle-\langle n_{i,\downarrow}\rangle\big)\ ,
     \label{magneticmoment}
\end{equation}
where $\mu_{B}$ is the Bohr magneton. Finally, the total magnetic moment per unit cell of the system is given by $M=\sum_{i}m_{i}$. 
We will present these magnitudes in section~\ref{sec:results}.

Finally, we employ the TB model to address the effect of hydrogen adsorption. A commonly used approach consists of substituting the H~adatom by a single vacancy defect, neglecting lattice relaxation. This  is motivated by the minimal lattice distortion produced by the H~adatom and the hybridization of the C orbital, which effectively removes the orbital from the low-energy spectrum. These two factors make it possible to assume that a vacancy in an undistorted lattice is equivalent to a H~adatom. Within this approximation, the vacancy is modeled as empty atomic site, removing the corresponding hopping terms in the electron Hamiltonian~\cite{Hidrogenated2010Rossier,rossier2008armchair}. For comparison, we  have also implemented a more realistic Anderson model, where the H~adatom is treated as an impurity~\cite{Anderson-Hubbard,imputiries_katsnelson,hydrogenatedmodel}. In this case, an additional term in the Hamiltonian is added to describe the impurity states and their interaction with the graphene lattice, given by
\begin{align}
  \mathcal{H}_\mathrm{imp}&=U_{H}h^{\dagger}_{\uparrow}h_{\uparrow}h^{\dagger}_{\downarrow}h_{\downarrow}+\sum_{\sigma}\epsilon_{H}h^{\dagger}_{\sigma}h_{\sigma}
  \nonumber \\
  &-t_{H}\sum_{\sigma}(c^{\dagger}_{0,\sigma}h_{\sigma}+H.c)\ , 
\end{align}
where $h^{\dagger}_{\sigma}$ and $h_{\sigma}$ are the creation and annihilation operators for an electron in the $1s$ orbital of the H~adatom, respectively. The C~atom bonded to the H~adatom is located at the atomic site $i=0$. We set the onsite energy $\epsilon_{H}=1.7\,$eV, the intra-atomic Coulomb repulsion $U_{H}=1.3\,$eV and the hopping energy between the C and the H~adatoms is set $t_{H}=5.2\,$eV. These values are obtained by fitting to our DFT calculations presented below.

\subsection{First-principles calculations}

We have performed first-principles calculations within the DFT framework using the SIESTA code with spin-polarization~\cite{Soler2002}. The crystal structures were optimized with the revised functional RPBE~\cite{Hammer1999}. The electron-ion interactions are modeled with norm-conserving nonlocal Troullier–Martins pseudopotentials described with a double-$\zeta$ singly polarized basis set. The energy cutoff is set to 400 Ry. The structure was relaxed by conjugate gradient  optimization until forces were smaller than 0.005 eV/\,\AA. This provides enough precision to obtain reliable strain properties. Periodic boundary conditions were applied along the longitudinal axis of the ZGNR, so we use sufficiently large supercell parameters ($20\,$\AA) in the perpendicular and transverse directions to prevent spurious interactions between adjacent nanoribbons. All C atoms at the edges were passivated by hydrogen. Finally, we have employed a Monkhorst-Pack scheme with $n \times 1\times 1$ $k$ points sampling of the Brillouin zone, where $n$ is set to 5000 $k$ points. This large number is essential to avoid a nonphysical magnetic response of the nanoribbon.

To obtain the strain properties we employ the following scheme: (i) We perform a full relaxed calculation with the ferromagnetic~(FM)  and antiferromagnetic~(AFM) guess configuration for each ZGNR. (ii) From the relaxed structure, we apply strain to the supercell along the longitudinal direction. (iii) To check how to address the transverse strain, we perform a relaxation 
on the transversal direction. This allows us to widen or narrow the nanoribbon depending on the compressive or tensile strain induced along the longitudinal direction. 
In this way, elastic properties such as the Poisson ratio can be obtained. In each case, we extract the total magnetic moments 
from the Mulliken spin-split populations for each orbital and atom. 

\section{Results}\label{sec:results}

In order to clarify the effect of point defects in strained ZGNRs, we analyze first the role of strain separately, both in FM and the AFM solutions.  This allows us to highlight the impact of atom adsorption in these systems. 

The maximum strain that graphene can withstand is approximately $25$\%, according to both theoretical~\cite{liu2007ab,zhao2009uniaxial} and experimental~\cite{2008science321,perez-garza2014} studies. In our study, we primarily consider strain values up to $20$\%, unless stated otherwise. It is worth mentioning that most experiments make use of local probe microscopes to induce strain in graphene. Previous numerical studies revealed breaking of valley degeneracy due to a nonuniform strain produced by out-of-plane deformation~\cite{Settnes2016,Zhai2018}. However, in this work we assume uniform strain and neglect valley polarization.

\subsection{Strain on pristine ZGNRs}

\begin{figure}[ht]
    \centering
    \includegraphics[width=1.00\columnwidth]{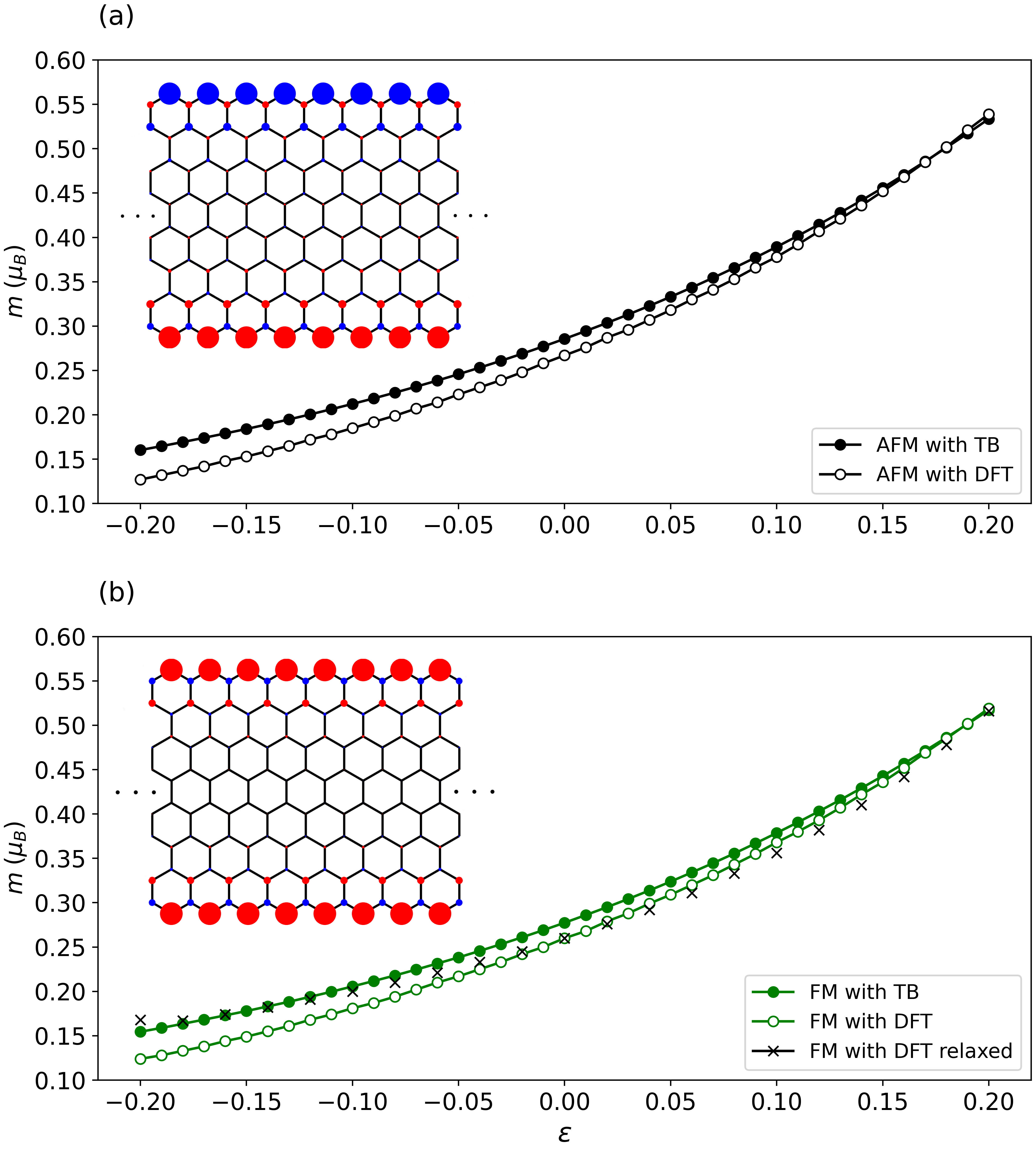}
    \caption{Spatial distribution of the magnetic moments in a 8-ZGNR (radii of the circles are proportional to the magnetic moment at each site, with red circles indicating positive values and blue circles negative ones) and value of the magnetic moment of the edge atoms as a function of strain in (a) the AFM configuration and (b) the FM configuration.}
    \label{fig2} 
\end{figure}

The Hubbard model predicts two possible magnetic solutions in a ZGNR depending on the relative orientation of spins at opposite edges. Due to the strong coupling between neighboring atoms, the spins within an edge are ferromagnetically coupled. However, edge-to-edge interaction is weaker and decays as $1/W^2$ with the ribbon width $W$, so two solutions can be explored. When the spins on one edge are antiparallel to those on the opposite edge, the solution is AFM, as depicted in Fig.~\ref{fig2}(a). If the spins at both edges are parallel, the solution is FM, as shown in Fig.~\ref{fig2}(b). An analysis of the total energy of both configurations reveals that the AFM solution is the ground state. However, since the energy difference with respect to the FM state is only a few meV even for narrow ribbons ($W=8$)  as those studied here~\cite{Pisani2007magneticribbons}, it is  easy to switch between the two configurations, for instance, by flipping the spins with a small magnetic field~\cite{Muñoz-Rojasmagnetoresistive}. Therefore, in view of this, we study both AFM and FM states.

In Fig.~\ref{fig2} we show the magnetic moment of the C atoms at the edges as a function of strain for the AFM and FM configurations. The edge atoms are the most relevant, since they provide the main contribution to the total magnetic moment. Positive (tensile) and negative (compressive) strain values are considered, following Eq.~\eqref{strain}. Note that compressive strain in graphene usually produces out-of-plane deformations like ripples or bending; for the sake of clarity, these effects are not considered in this work. The magnetic moments are computed using both the TB approach and DFT calculations. The edge magnetic moment is found to increase gradually and smoothly with strain, presenting the same trend in both magnetic configurations. It increases by $41$\% ($42$\%) with respect to the strain-free situation when the strain is $+0.10$ in the AFM (FM) configuration, according to DFT results. For an induced strain of $+0.20$, the magnetic moment is $102$\% ($101$\%) higher than the strain-free ZGNR in the AFM (FM) configuration. This effect  shows that the magnetic response of a strained ZGNR can be selectively enhanced. 

The excellent agreement between the TB and DFT results in Fig.~\ref{fig2} is remarkable. The difference is slightly higher in the case of compressive strain, but this condition is less common in actual experiments. For tensile strains, the maximum difference between both methods arise in strain-free samples and corresponds to a deviation in magnetic moment of $7.1$\% and $6.5$\% in Figs.~\ref{fig2}(a) and~\ref{fig2}(b), respectively. Furthermore, using DFT we examine the effect of the relaxation in the transverse direction due to the Poisson effect in the FM configuration [see black crosses in Fig.~\ref{fig2}(b)]. Interestingly, there is no significant difference between this result and the non-relaxed situation. For instance, the magnetic moment differs only by $6.1$\% between the relaxed and the TB non-relaxed cases when the strain is $+0.10$. This demonstrates the validity of our assumption that relaxation effects are negligible.

\begin{figure}[ht]
    \centering
    \includegraphics[width=0.9\columnwidth]{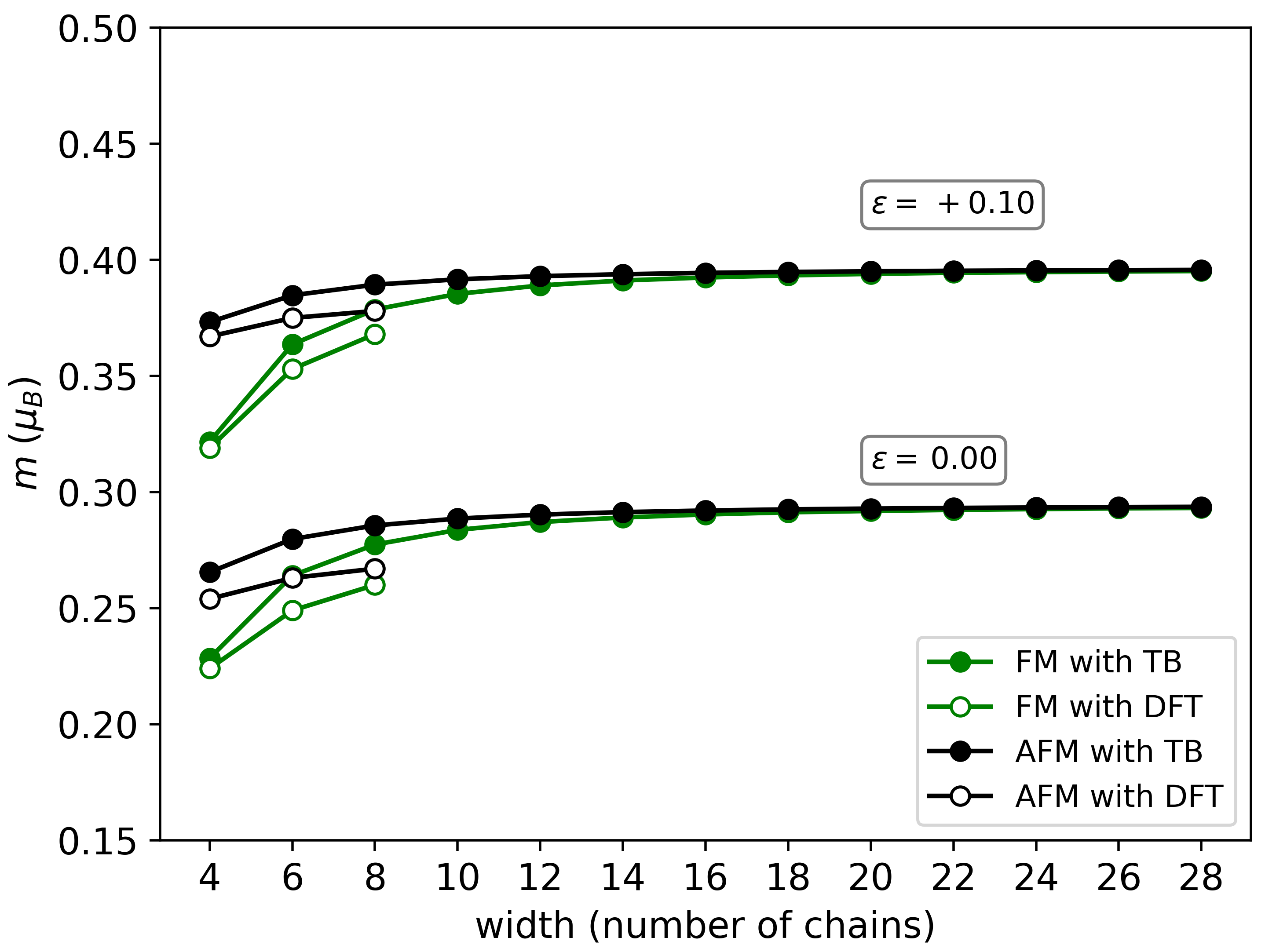}
    \caption{Magnetic moment of the edge atoms against the width  $W$ of the ZGNR.
    The graph includes data for two strain values, 0.00 and +0.10.}
    \label{fig3}
\end{figure}

We also study how the magnetic moment varies with the width of the ZGNR. Figure~\ref{fig3} depicts the edge magnetic moment plotted against the width expressed in number of zigzag chains $W$, for the strain-free and $+0.10$ strain cases. The AFM solution yields slightly higher values than the FM solution when the ZGNR is narrow. In both cases, the magnetic moment increases as the width increases. The TB calculations enable us to explore very wide ZGNRs, where the magnetic moment reaches saturation, with both the AFM and FM solutions converging to the same value. Additionally, the results obtained with DFT for narrow ZGNRs are also shown for comparison. However, scaling to large widths with this approach is time-consuming. In such cases, the TB model offers a much more efficient way to obtain reliable results in the case of wide ZGNRs.

\begin{figure}[ht]
    \centering
    \includegraphics[width=1.00\columnwidth]{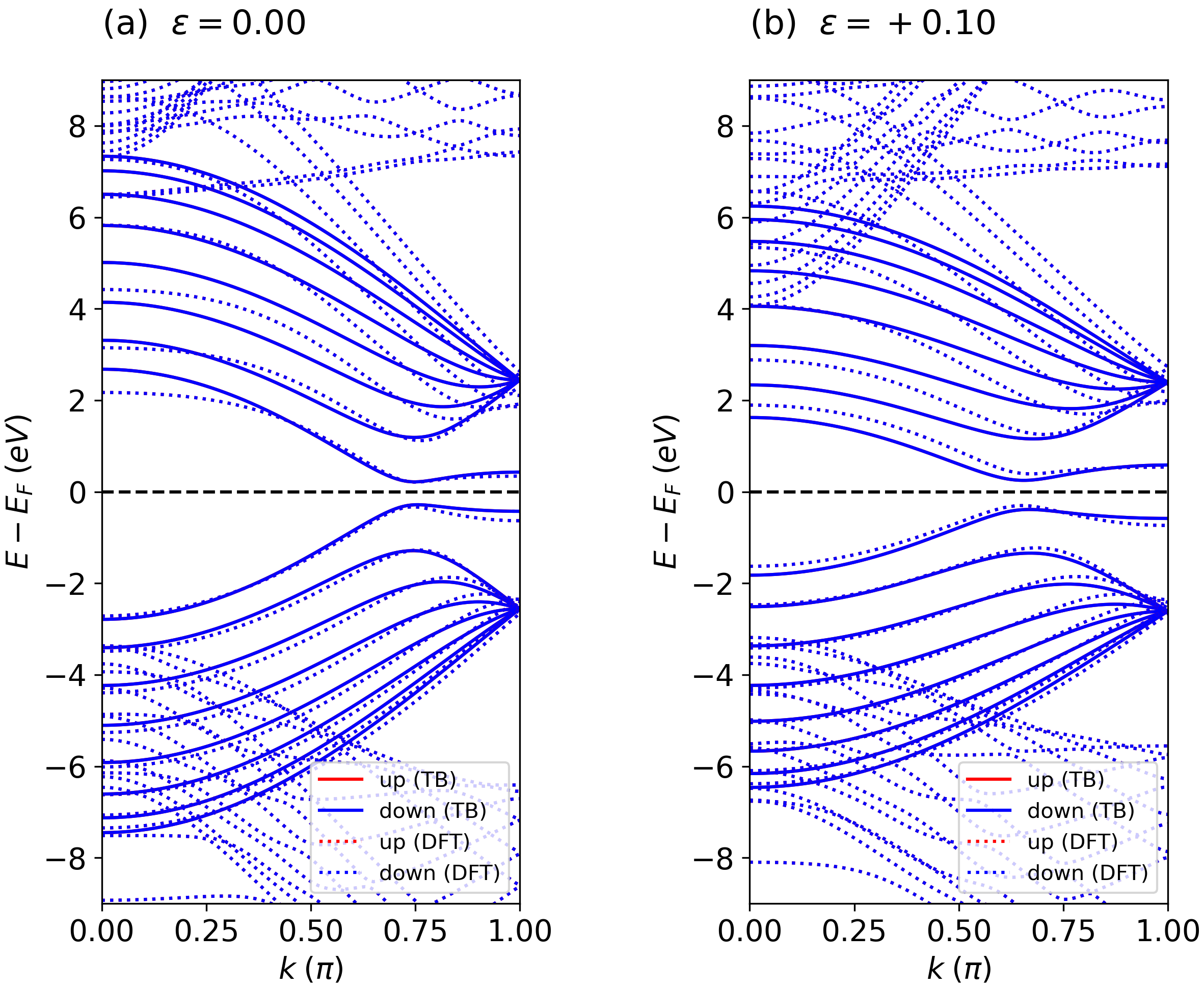}
    \caption{Band structure of the 8-ZGNR obtained with the TB (solid line) and DFT (dotted line) approaches in the AFM configuration for (a) strain-free and (b) strain of $+0.10$ cases.}
    \label{fig4}
\end{figure}

To elucidate the enhancement of the magnetic moment with strain observed in Fig.~\ref{fig2}, we analyze the band dispersions of the 8-ZGNR. Fig.~\ref{fig4} shows the band structures  for the AFM configuration with both the TB (solid line) and DFT (dotted line) approaches for two strain values. In the AFM configuration, up and down spin-polarized bands are degenerate, and a gap opens between the edge bands. The edge states open a gap, with the Fermi level in between. Figure~\ref{fig4}(a) corresponds to the strain-free case, while Fig.~\ref{fig4}(b) shows the behavior when the ZGNR is subjected to $+0.10$ strain. Strain does not change the overall aspect of the bands. However, two differences are apparent. Firstly, the bands become significantly flatter with strain, as expected, since the bandwidth is proportional to the hopping parameter and this decreases with increasing distance between C atoms. Secondly, edge states in the strained case are not only flatter, but they occupy a larger portion of the Brillouin zone. 
Since edge states are responsible for the magnetism, the enhancement of these states leads to an increase in the magnetic moment. 

\begin{figure}[ht]
    \centering
    \includegraphics[width=1.00\columnwidth]{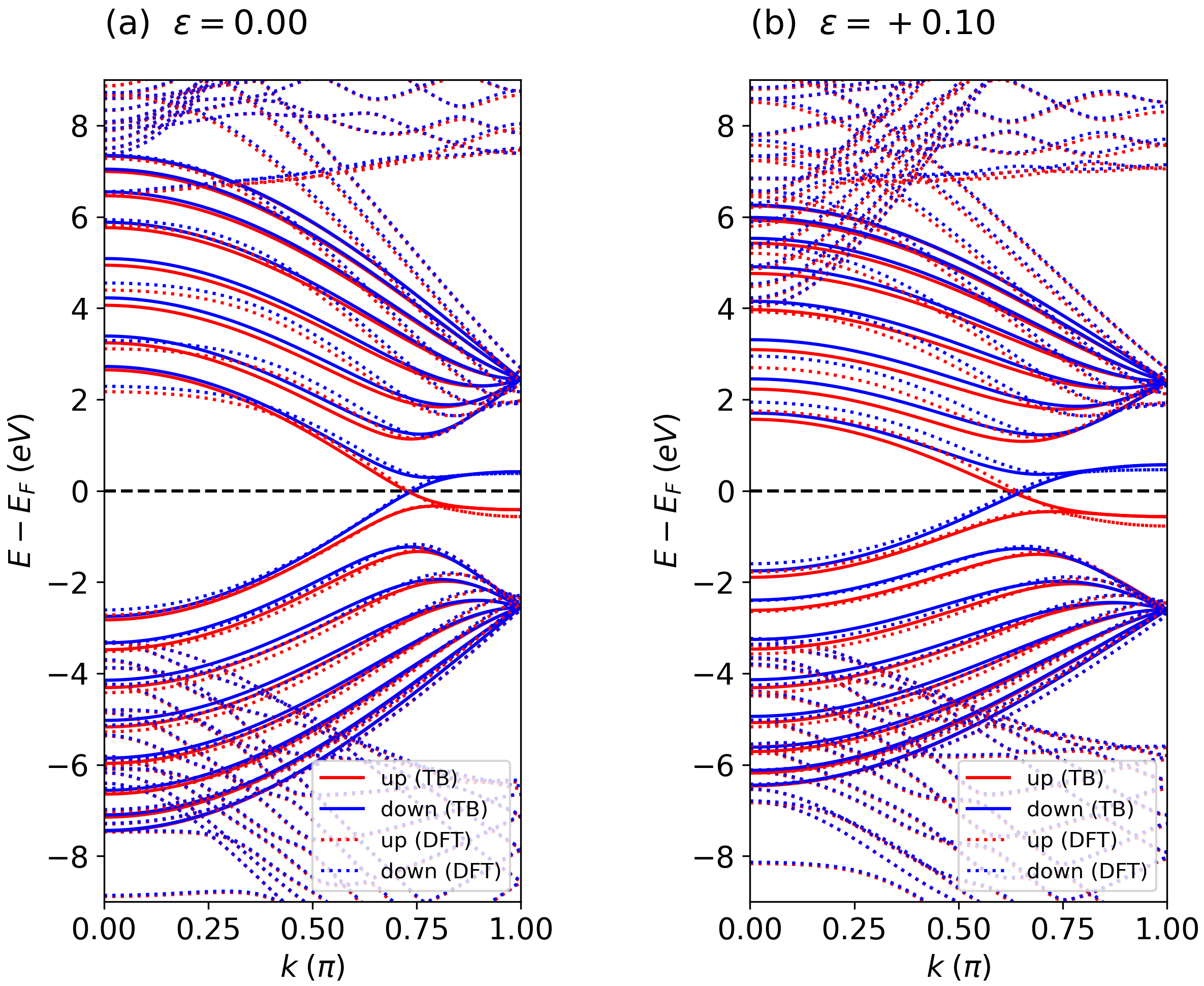}
    \caption{Band structures of the 8-ZGNR obtained with the TB (solid line) and DFT (dotted line) approaches in the FM configuration for (a) strain-free and (b) strain of $+0.10$ cases.}
    \label{fig5}
\end{figure}

We proceed analogously with the FM configuration of the 8-ZGNR.  
The band structure for the strain-free case is shown in Fig.~\ref{fig5}(a). In the FM solution, the degeneracy of the up and down spin bands is broken.
Now the system is gapless and the spin-up polarized edge band 
lies below the Fermi level, giving rise to a nonzero magnetic moment. The strained $+0.10$ case is shown in Fig.~\ref{fig5}(b). Once again, the overall aspect of the bands remains similar with applied strain, with the bands becoming flatter. 
Remarkably, the edge 
bands 
are more separated in the strained situation compared to the strain-free case. The Fermi level lies always symmetrically between 
the edge bands. 
Consequently, this separation, along with the fact that the flatter portions of the bands are larger, implies that more spin-up states are occupied 
than in the strain-free case, 
thereby increasing the net magnetic moment.

\subsection{H~adatoms in pristine ZGNRs}
 
We now consider the effect of H atoms adsorbed onto the ZGNR. For numerical calculations, an adatom is placed in a supercell that is three times wider the basic unit cell,
corresponding to a concentration of approximately $2$\%. In graphene, H~adatoms are well-known to induce a magnetic moment in the surrounding C~atoms of the opposite sublattice, resulting in a characteristic triangular or $\sqrt{3}\times\sqrt{3}$ pattern around the defect~\cite{yazyev2007}. Consequently, the magnetic moment of the system is modified. 

\begin{figure}[ht]
    \centering
    \includegraphics[width=0.75\columnwidth]{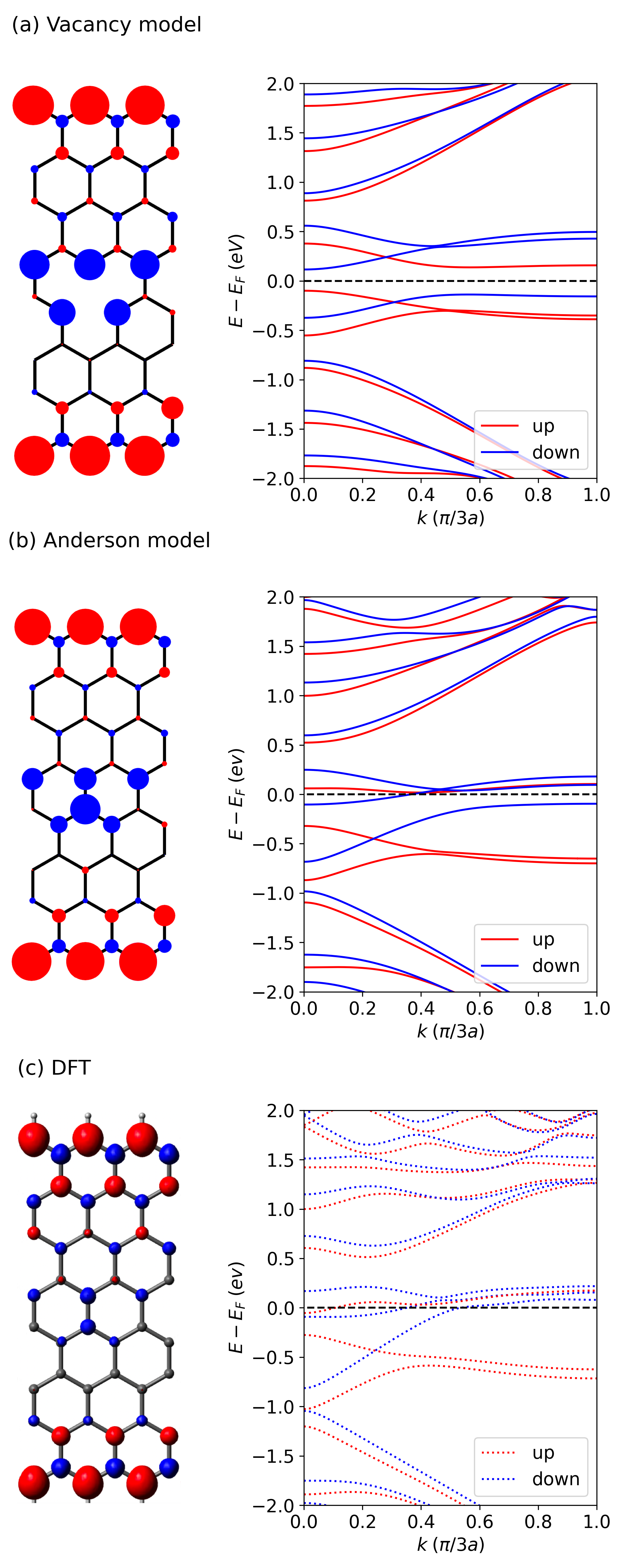}
    \caption{Spatial distribution of the magnetic moment with an H~adatom located at the 4th chain in the supercell (with FM configuration at the edges) and the corresponding band structure computed using (a) the vacancy-model approach (b) the Anderson-impurity model and (c) DFT calculations. Radii of the circles are proportional to the magnetic moment at each site, scaled by a smoothing factor to enhance visualization. Red circles indicate positive values and blue circles negative ones.}
    \label{fig6}
\end{figure}

We compute the adsorption of H~adatoms using three methods: the vacancy-model approach, the Anderson-impurity model and DFT calculations. Both the vacancy-model approach and the Anderson-impurity model are implemented within the TB framework, described in detail in Section~\ref{sec:Model}. Figure~\ref{fig6} presents the results obtained with these three methods, including the spatial distribution of the magnetic moment in the supercell and the corresponding band structure. Here the H~adatom is located at the central chain. All methods reveal a very similar electronic spatial distribution, characterized by a noticeable triangular pattern of magnetic moments with negative values at the C~atoms surrounding the H~adatom, specifically those in the sublattice opposite to the defect, as anticipated. In both the Anderson model [see Fig.~\ref{fig6}(b)] and the DFT results [see Fig.~\ref{fig6}(c)] the H~adatom also holds a prominent magnetic moment at the center of the triangle, differing from the empty space of the vacancy-model approach [Fig.~\ref{fig6}(a)]. In these cases, the magnetic moment from the bonded C~atom is hidden by the top H~adatom. However, we have confirmed that no magnetic moment arises in this atom, making it irrelevant to the overall magnetic structure.

The influence of edge states on the magnetic pattern is highly significant. We focus on the FM configuration between the edges, since its metallic behavior offers a more suitable platform for studying and manipulating defect states by mechanical deformation of the ZGNR, in contrast to the gapped AFM solution. The edge states notably affect the defect state around the H~adatom. To be specific, notice the absence of a significant magnetic moment at the lower vertex of the triangular magnetic pattern. In an infinite graphene sheet we would expect a negative magnetic moment at this atom, completing the well-known triangular magnetic pattern around the defect. This pattern is distorted due to its proximity to the lower edge of the ZGNR, where the electron exhibits spin-up polarization. Actually, the influence of the edge states is even more profound, determining the overall magnetic pattern. In contrast to the results for an isolated H~adatom in bulk graphene, where the magnetic pattern typically exhibits spin-up polarization~\cite{yazyev2007,Yazyev2010review,Anderson-Hubbard}, here the pattern is forced into a spin-down configuration. We have found that this spin-down state is energetically the most favorable solution, differing by approximately $0.27\,$eV/atom from a solution with a spin-up defect pattern. Thus, the defect state is antiferromagnetically coupled to the edge states of the ZGNR, resulting in a ground state with a spin-down polarization.

With respect to the band structure, all models provide the same low-energy bands for $\abs{E}<1\,$eV. For instance, in Fig.~\ref{fig6}(a), two spin-up bands appear below the Fermi level, merging and flattening as $k$ is closer to the boundary of the Brillouin zone. These bands correspond to the two spin-up edge states. Similarly, two equivalent spin-down bands are located above the Fermi level. Additionally, a single spin-down band near the Fermi level is related to the localized defect pattern, with its corresponding unoccupied counterpart. These six bands appear in all the models and methods chosen for this work. However, it is observed that the vacancy-model solution differs notably from DFT results, presenting an insulating gap and symmetric behavior around the Fermi level, in contrast to the metallic and asymmetric character of the DFT bands. In contrast, the bands obtained from the Anderson model exhibit a remarkable agreement with the DFT results. Then, we come to the conclusion that the common approach of simulating H~adatoms as a vacant atomic site in the graphene lattice is not reliable enough~\cite{rossier2008armchair,Hidrogenated2010Rossier,SRoche_spintransport}. As we have demonstrated, this approximation accurately describes the spatial distribution of magnetism, but fails to provide an accurate description of the band structure, most importantly, predicting an insulating behavior instead of metallic. In contrast, the Anderson-impurity model, which is computationally less demanding than DFT calculations, offers a more accurate description of the electronic bands.

\subsection{H~adatoms in strained ZGNRs}

\begin{figure*}[ht]
    \centering
    \includegraphics[width=2.08\columnwidth]{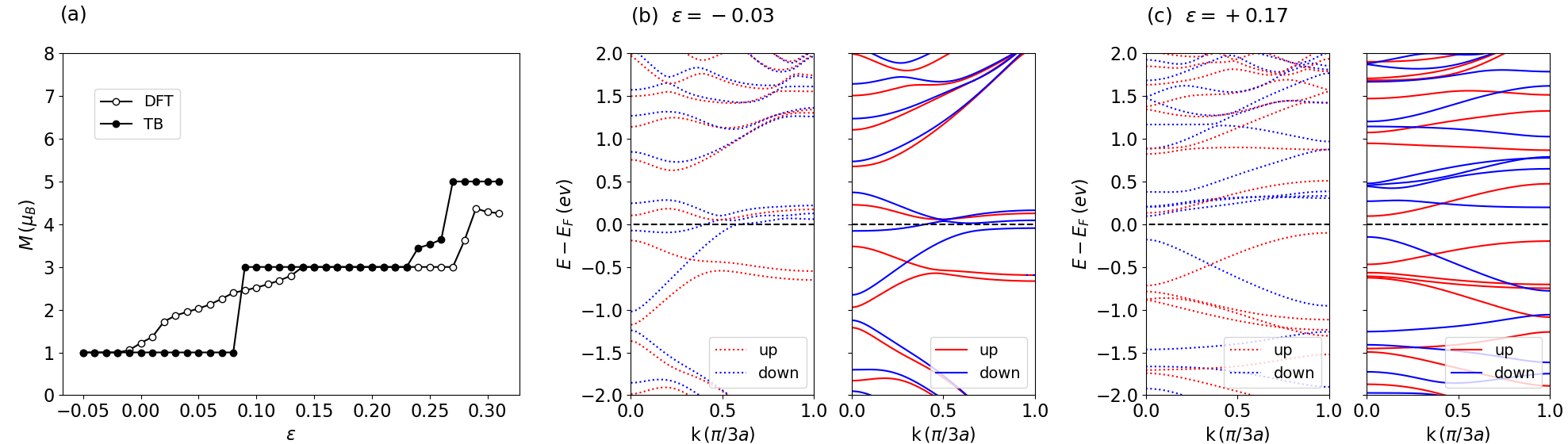}
    \caption{(a) Total magnetic moment as a function of strain in the H-adsorbed supercell. Band structures obtained from DFT (dashed line) and TB (solid line) methods for strain values of (b) -0.03 and (c) +0.17, corresponding to the first and second plateaus, respectively.}
    \label{fig7}
\end{figure*}

Finally, we investigate the magnetic behavior of the H-adsorbed ZGNR in combination with strain. Due to the spatial symmetry breaking introduced by the adatom, it is more appropriate here to analyze the total magnetic moment $M$ rather than focusing only on the edge magnetic moment, as we presented in Fig.~\ref{fig2}. In Fig.~\ref{fig7}(a) we show the total magnetic moment as a function of strain, from $-0.05$ to $+0.30$. The reason for this choice is to prevent out-of-plane deformations caused by high compressive strains, while providing an ample window for tensile strain. The figure shows that the magnetic moment increases under strain. Interestingly, unlike the smooth growth observed in Fig.~\ref{fig2}, the magnetic moment exhibits a stepped behavior, characterized by plateaus where it remains constant. Remarkably, the magnetic moment has exactly odd integer values at these plateaus, with three distinct steps at  $M(\mu_{B})=1,3$ and $5$. This behavior is observed in both the Anderson model and DFT calculations. However, the widths of the plateaus obtained with both approaches are slightly different, and the transitions between plateaus are abrupt in the Anderson model, in contrast to the smoother transitions observed in DFT results. Nevertheless, both methods predict the existence of plateaus; they arise from the same underlying mechanisms. 

To analyze the origin of the steps in the magnetic moment as a function of strain, we present the band structure for strain values of $-0.03$ [see Fig.~\ref{fig7}(b)] and $+0.17$ [see Fig.~\ref{fig7}(c)], which correspond to the first and second plateaus, respectively. Again, the good agreement between the band structure obtained from the Anderson model and the DFT calculations is appartent. At strain $-0.03$, the bands exhibit half-metallic behavior. A half-metal is characterized by being an insulator for one spin orientation, while remaining metallic for the other~\cite{half-metal}. As observed in Fig.~\ref{fig7}(b), for spin-up polarization, the bands are filled and separated by a bandgap from the unoccupied bands. In contrast, the spin-down bands are partially occupied, showing a metallic behavior for this spin polarization. A central characteristic of the half-metallic systems is the quantized value of  magnetization~\cite{half-metal}, which in our system manifests as the $M=\mu_{B}$ plateau. This arises because the system contains an integer number of $N_{\uparrow}$ spin-up electrons due to the filled bands, as well as an integer total number of electrons $N=N_{\uparrow}+N_{\downarrow}$ in the supercell, where $N_\downarrow$ is the number of spin-down electrons. Consequently, there must be an integer number $N_{\downarrow}$ of spin-down electrons even if the corresponding bands are partially filled. Comparing the DFT bands from Fig.~\ref{fig7}(b) with the DFT strain-free adatom case depicted in Fig.~\ref{fig6}(c), we observe that the bands are very similar. The key difference is that in the unstrained DFT case the spin-up defect band crosses the Fermi level near $k=0$, departing from the half-metallic behavior. Therefore, the system undergoes a transition from half-metallic to being metallic for both spins, which explains why the magnetic moment deviates from the plateau at zero strain in Fig.~\ref{fig7}(a) for the DFT calculation. In contrast, in the strain-free TB results from Fig.~\ref{fig6}(b), the spin-up band does not cross the Fermi level, and the system remains half-metallic. This explains the persistence of the plateau at positive strain values in Fig.~\ref{fig7}(a), unlike the behavior observed with DFT calculations.

Figure~\ref{fig7}(c) depicts the bands for the strained ZGNR with $\epsilon=+0.17$, for which an integer magnetic moment is found both in the DFT and the TB model. For this strain the system is no longer half-metallic since the band structures in both approaches display a bandgap for the two spins. The origin of this second quantized plateau differs from the previous case and stems from the integer number of occupied bands for both spin polarizations: the imbalance between the number of filled spin-up and spin-down bands gives the integer value of the magnetic moment. The same happens for the  $M=5\mu_{B}$ plateau.  However, DFT calculations do not reach the plateau predicted by the TB approach for large strains, probably due to the substantial distortion effects caused by the huge strain values required to reach that plateau. 

The previous results present intriguing possibilities for the systems studied in this work. Firstly, half-metallic materials are highly desired for spintronic applications as a means to generate completely spin-polarized currents. The realization of half-metallicity in ZGNRs has already been predicted in different setups, such as the use of electric fields~\cite{son2006halfmetallic}, edge modification~\cite{edgemodified-halfmetal}, substitutional doping~\cite{Pt-impurities,BN-impurities} or magnetic atom adsorption~\cite{half-metaladatoms}, typically transition metals. However, these situations tend to be experimentally more complex than the adsorption of H~adatoms. Secondly, the appearance of quantized magnetic moments may hold potential for future magnetic applications, given that the required conditions are not exceedingly challenging, namely the use of H~adatoms, which are among the most common dopants, and zero (or slightly negative) strain values to reach the first plateau.

Finally, it is worth mentioning that our choice of the 4th chain position of the H~adatom in this work is only motivated for a better visualization of the magnetic pattern. We have also explored other configurations with the H~adatom in different chains. We have found that the results are consistent with those presented here, including the magnetic pattern, half-metallic behavior, and quantized plateaus in the magnetic moment. Only when the H~adatom is located just at the edges, and depending on its sublattice, it strongly interacts with the edge states, but this situation is in fact an edge modification, outside the scope of our work. 

\section{Conclusions}

In summary, we have investigated the effects of strain and H~adatom adsorption on the magnetic properties of ZGNRs using a combination of TB models and DFT methods.  A longitudinal strain applied along the ZGNRs makes the magnetic moments in both FM and AFM configurations increase progressively with strain, achieving a significant growth rate. This enhancement in magnetic moment is attributed to strain-induced modifications in the band structure, and has a smooth behavior. Our TB model shows excellent agreement with DFT calculations and enabled us to study much larger systems with reduced computational effort compared to DFT calculations. We have explored the effects of H~adatoms on ZGNRs with different computational approaches. We found that modeling H~adatoms as vacancies do not accurately capture the band structure, leading to misleading conclusions. The Anderson impurity model, however, yields excellent agreement with the DFT-derived band structure. The magnetic configuration of H adatoms is largely influenced by the edges of the ZGNR, where the ground state exhibits AFM coupling between the defect and the edges. Interestingly, H~adatoms induce a half-metallic character in ZGNRs. When combined with strain, it induces robust, quantized magnetic moments, characterized by distinct plateaus of integer values with varying strain. While the first plateau induced by smooth strain arises from the half-metallic character, subsequent plateaus observed at higher strains emerge due to the transition to a gapped state. These findings offer valuable guidelines for the manipulation of magnetism in ZGNRs and advance our capabilities for tuning magnetic properties in two-dimensional materials.

\acknowledgments

This work was supported by the Spanish Ministry of Science and Innovation (Grant PID2022-136285NB-C31 and PID2022-136636OB-I00). We also acknowledge the support from the “(MAD2D-CM)-UCM” project funded by Comunidad de Madrid, by the Recovery, Transformation and Resilience Plan, and by NextGenerationEU from the European Union. We also acknowledge the computational resources and assistance provided by the Centro de Computación de Alto Rendimiento CCAR-UNED and Servicio de SuperComputación SSC-UCLM.




%

\end{document}